# Tunable phase transitions in half-Heusler TbPtBi compound

Pratik D. Patel[a, b*], Akariti Sharma[a], Bharathiganesh Devanarayanan[a, c], Paramita Dutta[a], Navinder Singh[a]

[a]Theoretical Physics Division, Physical Research Laboratory, Navrangpura, Ahmedabad, India-380009

[b]M.B. Patel Science college, Anand-388001

[c]Indian Institute of Technology, Gandhinagar, Palaj, Gujarat, India - 382355

[*]Corresponding author: pratikphysics05@gmail.com

**Abstract**

We report various phase transitions in half-Heusler TbPtBi compound using Density Functional Theory (DFT). Specifically, the inclusion of spin-orbit coupling (SOC) leads to the band inversion resulting in the transition from the metallic to the topological semimetallic phase. However, in the presence of SOC, there is a phase transition from the topological semimetal to the trivial semimetal when the material is subjected to compressive strain ($\leftarrow 7\%$). Subsequently, under the further increase of compressive strain($\geq -7\%$), we find an opening of a direct band gap at the point, driving the system from the trivial semimetallic to a semiconducting state with changes in the sequence of the bands. In the absence of SOC, only the transition from the metallic to the semiconducting phase is noticed. Under tensile strain, the TbPtBi compound maintains its phase as in the unstrained condition but with an increase in the hole pocket at the Fermi level, both in the absence and presence of SOC. These tunable phase transitions (especially as a fraction of strain) make this compound very promising for application in various quantum devices, such as highly sensitive strain gauges.

## 1. Introduction

During the last few decades, half-Heusler (HH) compounds have been explored as one of the most promising candidates for spintronics and thermoelectric devices due to their high spin-polarization and profound thermoelectric efficiency [1-5]. These materials have the potential for applications in high-temperature sustainable thermoelectric devices due to their high thermoelectric coefficients and thermodynamical and mechanical stabilities [3-5]. The simple chemical formula XYZ defines the elemental compositions of HH compounds. Here X and Y are transition metal elements, and Z can be $p$ or $d$ block metal elements [5]. A single crystal of TbPtBi of dimensions 1.5x1.5x1.5mm$^3$ can be grown from Bi flux by the crystal growth technique [6]. Experiments on the powdered single crystal confirmed the MgAgAs-type crystal structure with a lattice parameter of 6.665 Å [6].TbPtBi single crystals exhibit semi-metallic behaviour with a broad peak determined within the ohmic resistance activity around 225K [7], as verified by Zhu et al. through an experiment.

This class of compounds possesses highly tunable electronic properties and the possibility of topological phases like topological insulators, topological semimetals, and Weyl semimetals [8-13]. Many ternary HH compounds are close to the boundary between the trivial and topological phases and can serve as a fantastic platform to explore the topological phase transitions [14-17,33-36]. These topological phases in some HH compounds have been found in the presence of spin-orbit coupling (SOC). The presence of SOC may lead to the band inversion phenomena in the band structures, indicating the transitions from the trivial to the topological phases or vice versa in those HH compounds [18-19]. In more detail, the electronic properties of HH compounds of the form XPtBi (X=Sc, Pt, and Y) have been studied by Goyal *et al*. [16]. They have found that these compounds show the trivial state with zero band gap in the absence of SOC, while the topological semimetallic phase is obtained in the presence of SOC. Interestingly, in addition to the topological phases, HH compounds like YPtBi and GdPtBi can show superconducting ground states at low temperatures, as explored by Yang *et al*. through their DFT calculations [17-19].HH compounds can also host Dirac states, as observed via angle-resolved photoemission spectroscopy (ARPES) measurements in the YPtBi compound [20]. The electronic structures of these compounds can be modulated by external effects like doping [21].

TbPtBi, one of the HH compounds, exhibits semimetallic behavior with a broad peak in the resistivity profile around 225K, experimentally observed by Zhu *et al*.[7].The ground state of TbPtBi is antiferromagnetic (AFM) below the Neel temperature ($T_N$) of 3.4K, as reported previously via magnetic measurements [22]. Moreover, AFM states of TbPtBi can show a significant anomalous Hall effect with a wide range of hall angles (0.68 to 0.76) reported recently [6,22]. Despite the prevalence of experimental studies on HH, there appears to be no theoretical study to date on the possibility of getting and tuning topological phases in TbPtBi to the best of our knowledge. How the presence of SOC affects the ground state of TbPtBi is not still explored.

Motivated by this, we study the TbPtBi compound of the HH family as this compound is less explored compared to the other materials of the HH class. We investigate various phase transitions subjected to an externally applied strain, both compressive and tensile, in the absence and presence of SOC. We find that TbPtBi shows a phase transition from the trivial to the topological phase with the inclusion of SOC and show that a small variation in the lattice constant caused by the strain can transform such material back to the trivial state again. However, depending on the degree of strain, it can show either a semimetallic or semiconducting phase when the strain is compressive. Under tensile strain, TbPtBi remains metallic with increasing hole pockets in the absence of SOC, while it shows the semimetallic phase with increasing hole pockets associated with decreasing electron pockets in the presence of SOC.

## 2. Computational details

The DFT calculations of the present work are done using the QUANTUM ESPRESSO code, which is widely used for geometry relaxation to find the minimum potential energy and is reliable for computational accuracy and reduced calculation time [23]. To obtain the energy of the TbPtBi compound accurately, the generalized gradient approximation (GGA) is employed to treat the exchange-correlation energy [24]. The plane wave basis set with an energy of 0.02 Ry is utilized to extend the Kohn-Sham orbitals. A force of 0.01 eV/Å is used to relax the unstrained and strained structure of the TbPtBi compound. The electronic properties of TbPtBi are investigated with and with out the SOC with scalar relativistic (SR) and fully relativistic (FR) Perdew-Burke-Ernzerhof (PBE) type pseudo potentials. TbPtBi crystal structure is relaxed with 60 Ry kinetic energy limit and 12×12×12 k-point Monkhorst grid. This compound is characterized by the face-centered cubic MgAgAs type structure, as confirmed experimentally [21]. Based on that, we consider Tb, Pt, and Bi atoms located at (0.5,0.5,0.5), (0.25,0.25,0.25) and (0.0,0.0,0.0) atomic sites, respectively, to relax the unit-cell of TbPtBi.

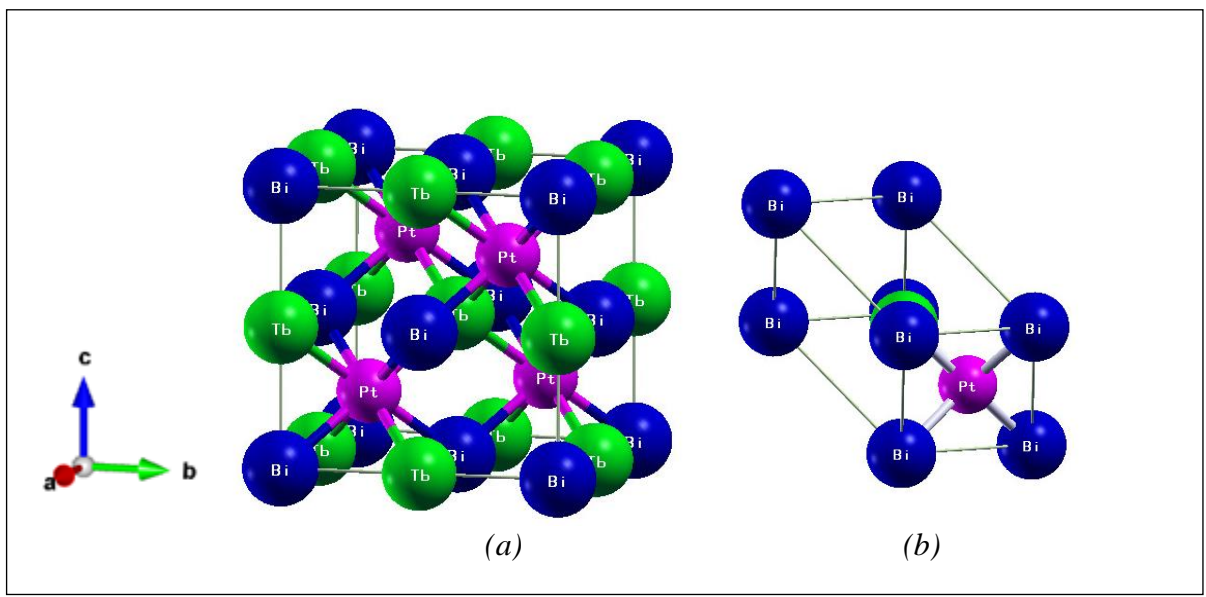

*Figure 1.* Optimized (a) conventional and (b) primitive unit cell of TbPtBi.

After relaxing the unit cell structure into AFM configuration within the GGA exchange correlation functional scheme, the minimum ground state energy of the primitive and conventional unit cell is calculated. Due to its low total energy of −842.2641351Ry and lattice parameter of 6.755Å, the AFM configuration is found to be the most stable and favourable for the TbPtBi structure. These results agree very well with the previously reported lattice constants and AFM states of these compounds ($T_N = 3.4$K) based on experiments [23]. Strain-dependent electronic band structures have been estimated using that relaxed lattice parameters along the high symmetry direction of W-L-Γ-X within the first Brillouin zone [25].

## 3. Anomalous electronic properties:

In this section, we study the various anomalous electronic phases of TbPtBi, including the effects of SOC and strain individually and simultaneously.

### 3(a). Effect of SOC

We start by describing the band structures to study the phase transitions induced by SOC. The electronic properties of the TbPtBi compound are investigated with relaxed structure and lattice parameters. At equilibrium lattice constants (in an unstrained environment), the electronic band structure of the compound is illustrated in Fig.2 (a) and Fig.2 (b) in the absence and presence of SOC. In Fig.2(a), the conduction band crosses the valence band above the Fermi level at the Γ point of the Brillouin zone. This results in the appearance of the big hole pocket at the Fermi level between A and B points, as shown in Fig.2(a). This metallic phase turns out to be affected when the SOC is included. The strong SOC leads to band splitting. A large shift occurs among the valence bands and the ordering gets affected. The $\Gamma_6$ band gets positioned below the $\Gamma_8$ band giving rise to the band inversion. More specifically, the four-fold degenerate state at $\Gamma_8$ band in Fig.2 (b) has $p$-type orbitals, whereas the $s$-type state is present in $\Gamma_6$ band. The lower energy of $s$-type orbital (-1.18eV) than the $p$-type orbital (0.12eV) caused due to the strong SOC describes the band-inversion.

Because of the heavy elements present in the compound, the effective mass at the Fermi level is increased, reducing the energy of the *s*-type orbitals as a consequence [25-27]. This change in the band order, i.e. *p*-type states in the $\Gamma_8$ band above the *s*-type orbitals, describes the band-inversion between $\Gamma_8$ and $\Gamma_6$ bands of TbPtBi, and it is similar to the band inversions reported in the HH GdPtBi and LuPtBi compounds [25,27]. It indicates the transition from trivial metallic to the topological semimetallic phase of the TbPtBi compound.

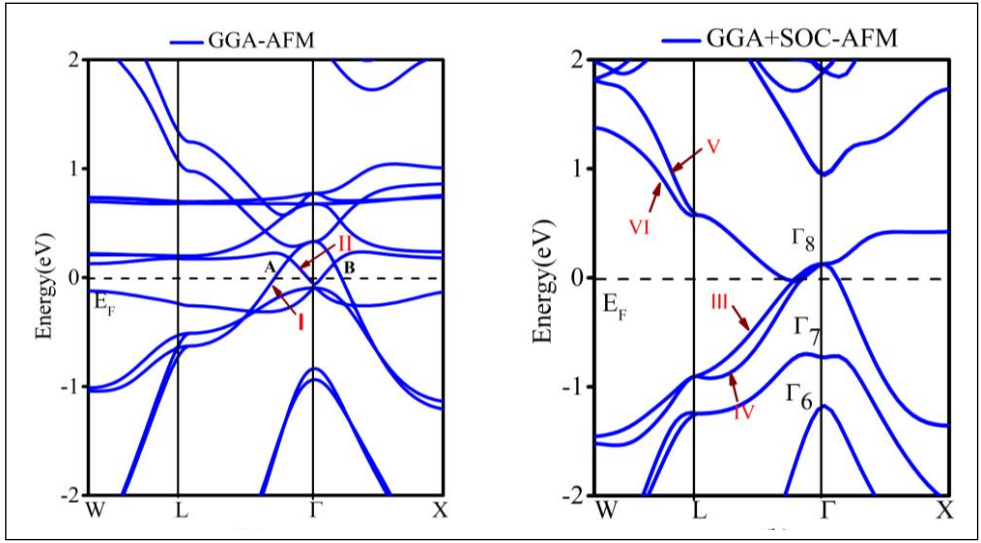

*Figure 2:* Bulk band structure of TbPtBi compound (a) without and (b) with SOC at equilibrium condition. The Fermi energy is denoted by the dashed lines.

The energy difference between the $\Gamma_8$ and $\Gamma_6$ bands defines the band inversion strength, which is denoted by: $\Delta E = E_{\Gamma_8} - E_{\Gamma_6}$. Positive and negative values of ΔE confirm the topological non-trivial and trivial nature of the compounds [17, 25-28]. The calculated band inversion strength for TbPtBi compound is 1.23 eV. The cubic symmetry of this compound leads to a topologically nontrivial semimetal that crosses the Fermi energy by $\Gamma_8$ bands with quadruple degeneracy. When SOC is included, the ordering of the bands' sequence from high to low energy is $\Gamma_8 - \Gamma_7 - \Gamma_6$ (see Fig.2(b)).

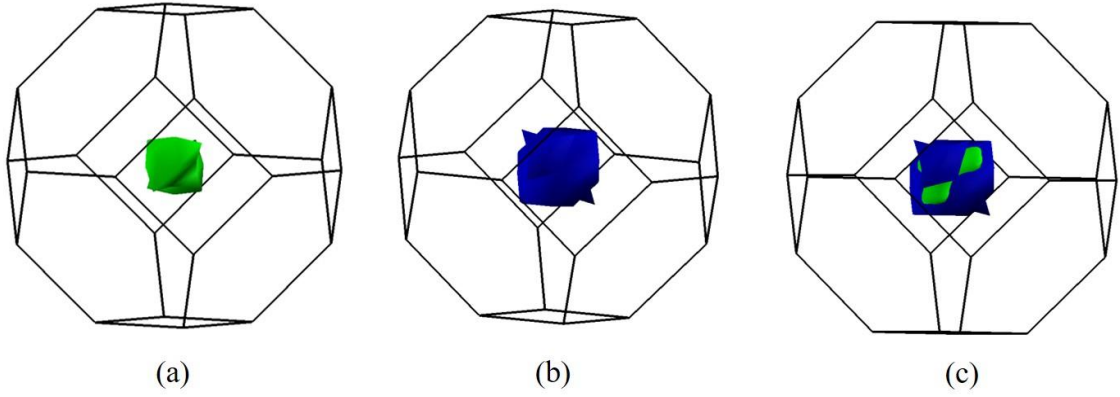

(a) (b) (c)

*Figure 3:* Fermi surface for the cubic structure of TbPtBi compound without SOC. The green (a) and blue (b) colored regions at Γ point (in the band structure in Fig-2(a)) confirms the presence of electron pockets due to the crossing of bands 'I' and 'II' in the band structure, while (c) merged Fermi surface sheet of these two fermi surfaces is presented.

Next, we check the Fermi surfaces for the electronic band structures of the TbPtBi compound and show in Figures 3(a-c) and 4(a-e) in the absence and presence of SOC, respectively. To construct the Fermi surface, we choose the electron and hole pockets formed around the Fermi level in Fig.2(a) and (b). In the absence of SOC, the heavy electron pocket at Γ point is formed by the 'I' and 'II' bands in the electronic band structure, as shown in Fig-2(a). The formation of the electron pocket is confirmed in Fig. 3(a) (the green colored region). In Fig.2(b), the large hole pocket at the Γ point in the presence of SOC is primarily contributed by the bands 'III' and 'IV' crossing from the valance band to the conduction band, whereas a smaller electron pocket in the Γ-L direction in the electronic band structure is dominated by the bands 'V' and 'VI' crossing from conduction band to valence band. Similar contributions of the bands of ScPtBi compound to the Fermi surface construction are shown by Majumdar *et al*.[27]. In Fig.4, the two square blocks in (a) and (b) represent the hole pockets in the Fermi surfaces, which are formed by the crossing of two bands (III and IV) in the band structure of Fig.2(b). In contrast, the Fermi surfaces in (c) and (d) with nested characteristics represent the lighter electron pockets generated by the VI and V bands of Fig.2 (b). The merged distribution of a Fermi surface with a hole and an electron pocket is shown in Fig.4(e).

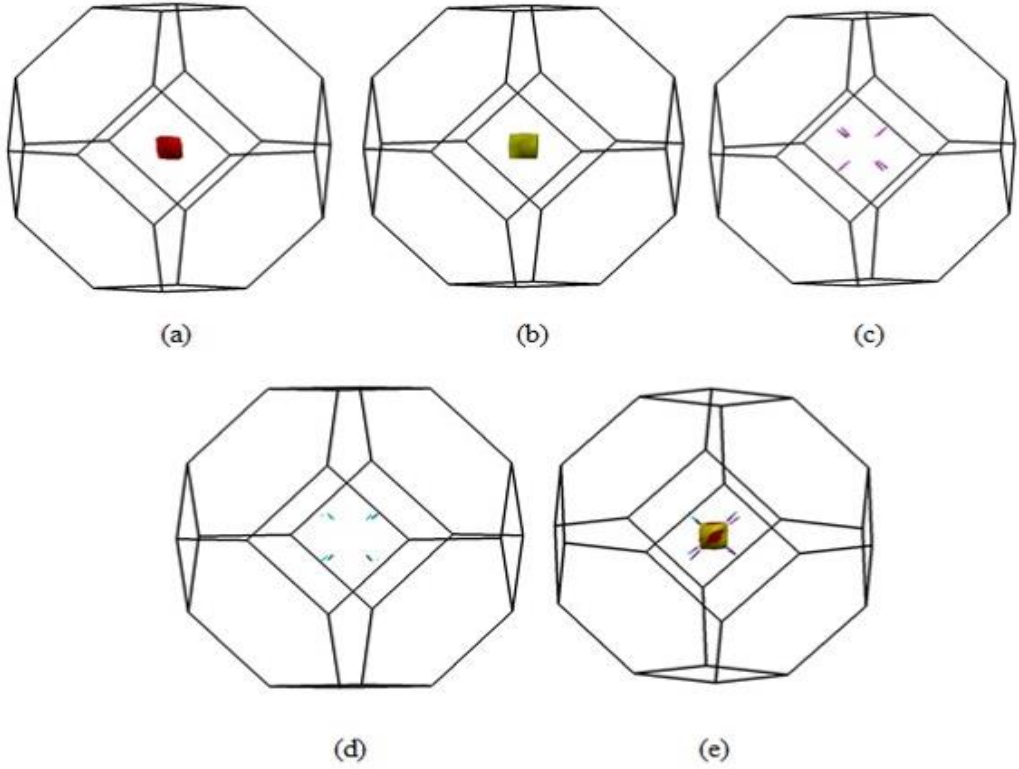

*Figure 4:* Fermi surface distributions in the presence of SOC. The red (a) and green (b) sheet confirm the existence of a large hole pocket at Γ point due to the bands crossing 'III' and 'IV', while (c) magenta and cyan color with nesting features confirms the presence of lighter electron pockets due to bands 'V' and 'VI'.

### 3(b). Effect of strain

Next, we investigate what happens when applying strain, both compressive and tensile. In this subsection, we show the strain-dependent electronic band structures without SOC in Fig.5(a-b). The band structure without SOC (in Fig.2(a)) confirms the metallic nature with a large hole pocket at Γ point. When compressive strain is applied, as shown in Fig.5(a), the size of the hole pocket in the band structure decreases. It continues till the compressive strain is −4%. We further increase the strain, and we notice that at −6% of compressive strain, the narrow direct band gap of 0.16 eV opens due to the shifting of the valence band edge to the conduction band side. In Fig.5(a), the indirect band gap of 0.32 eV is seen in the direction of W-Γ symmetric point in the band structure at -8% of compressive strain. So, there is a metallic to semiconducting phase transition in the band structure under the compressive strain of −4% to −6%. This phase transition is sensitive to the degree of the compressive strain.

Instead of the compressive strain, if we apply tensile strain, the TbPtBi compound remains in the metallic phase as it is in an unstrained condition irrespective of the degree of the tensile strain. We systematically present all the results under tensile strain in Fig.5(b). Note that we continue increasing the tensile strain until 8%. This type of behavior was previously observed by Xiao *et al.* In the LuPtBi compound[25]. The only change is: the size of the hole pocket increases with the increase of tensile strain in the TbPtBi compound.

### 3(c). Effect of SOC and strain

Now, we move on to the scenario when both SOC and strain are present. We show all the results for the compressive and tensile strain in Fig. 6. In the previous subsection, it is confirmed that there is no band gap in unstrained conditions in the presence of SOC (see Fig.2(b)). In the presence of SOC, it shows a topological semimetallic phase, as depicted in Fig.6(a). When we apply the compressive strain in the presence of SOC, we observe two topological phase transitions in the band structures: (i) from the topological semi metallic phase to the trivial semimetallic phase at −4% of compressive strain. (ii) from the trivial semimetal to the semiconductor when the compressive strain reaches -8%. When the compressive strain is increased from -4% to -6%, the band order ($\Gamma_8 - \Gamma_6$) is reversed to a regular band order ($\Gamma_6 - \Gamma_8$) [29,30]. A band gap of 0.59 eV is opened. It is much larger than the band gap in the absence of SOC under the compressive strain. The band structure of TbPtBi varies with increasing compressive strain (hydrostatic pressure), and therefore the energy gap changes from zero to 0.59 eV as a result of the applied pressure reducing the bond length; however, it will increase the bond energy of TbPtBi. Because the band gap is strongly influenced by the crystal field, it increases in size as total energy increases. The phase of the TbPtBi compound is highly tunable under compressive strain. This enhances the potential of the TbPtBi compounds very much in the tunable devices regarding practical application.

In Fig. 6(b), we present the results for tensile strain. Note that before the tensile the compound was in the topological semimetallic phase. As soon as we apply tensile strain, the hole increases in size. With the increase in the tensile strain, the size of the hole pocket increases from 0.18eV (at 2%) to 0.32eV (at 8%). In between 2% − 8% often tensile strain, the band inversion feature is characterized by the negative band inversion strength in the band structure, and it retains the semi metallic nature. Finally, we summarize all the possible phases in Table.1.

**Table 1:** Various phases of TbPtBi under various conditions of strain and SOC

| External constraint | Without any strain | Under compressive strain | | Under tensile strain |
|---|---|---|---|---|
| | | < -5% | ≥-5% | |
| Without SOC | Metal | Metal | Semiconductor | Metallic with increasing rate of hole pockets |
| With SOC | Topological semimetal | Trivial semimetal | Semiconductor | Semi-metallic with increasing hole pockets and decreasing electron pockets |

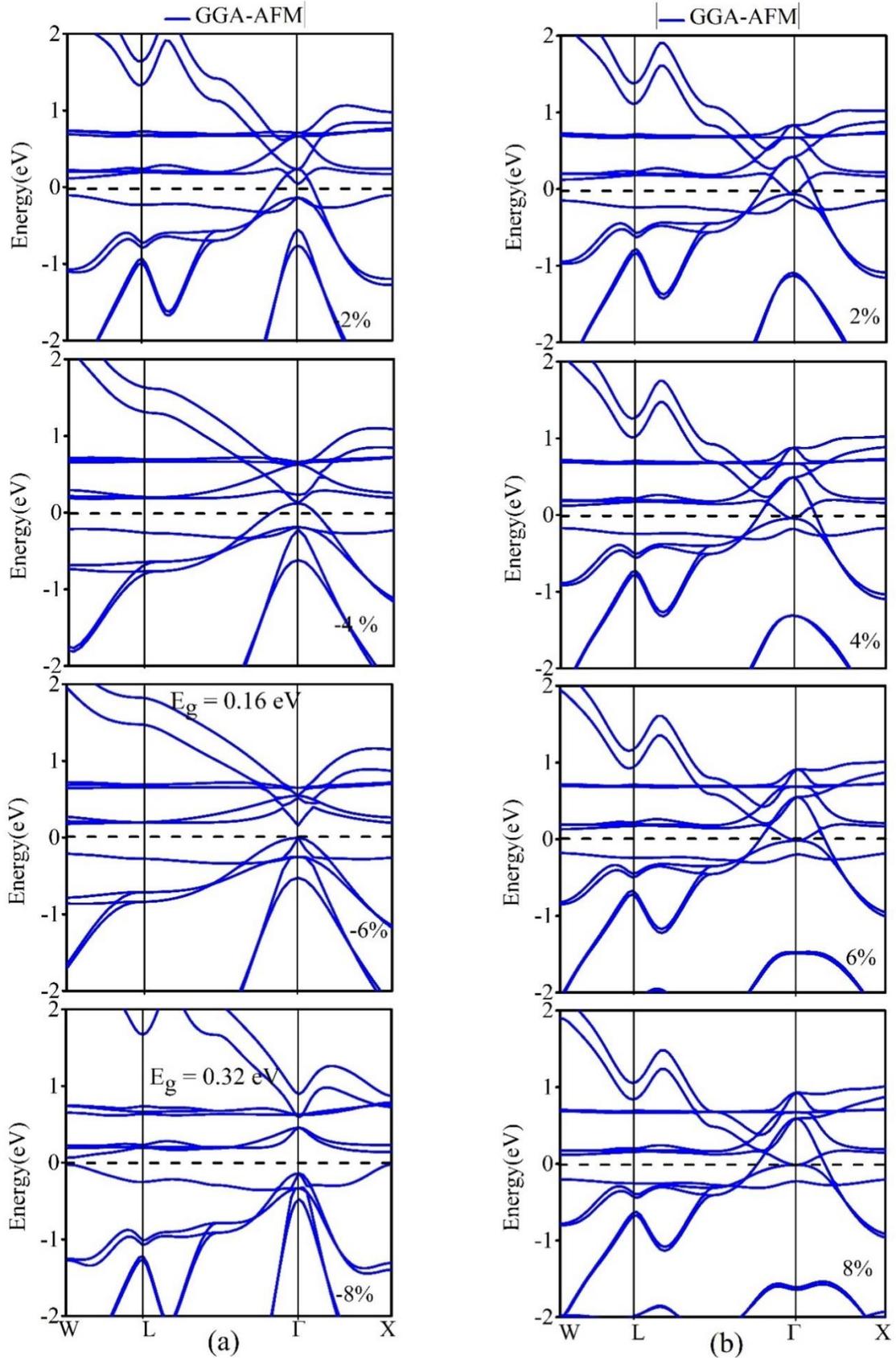

*Figure 5:* Band structures of TbPtBi under (a) compressive strain (b) tensile strain without SOC.

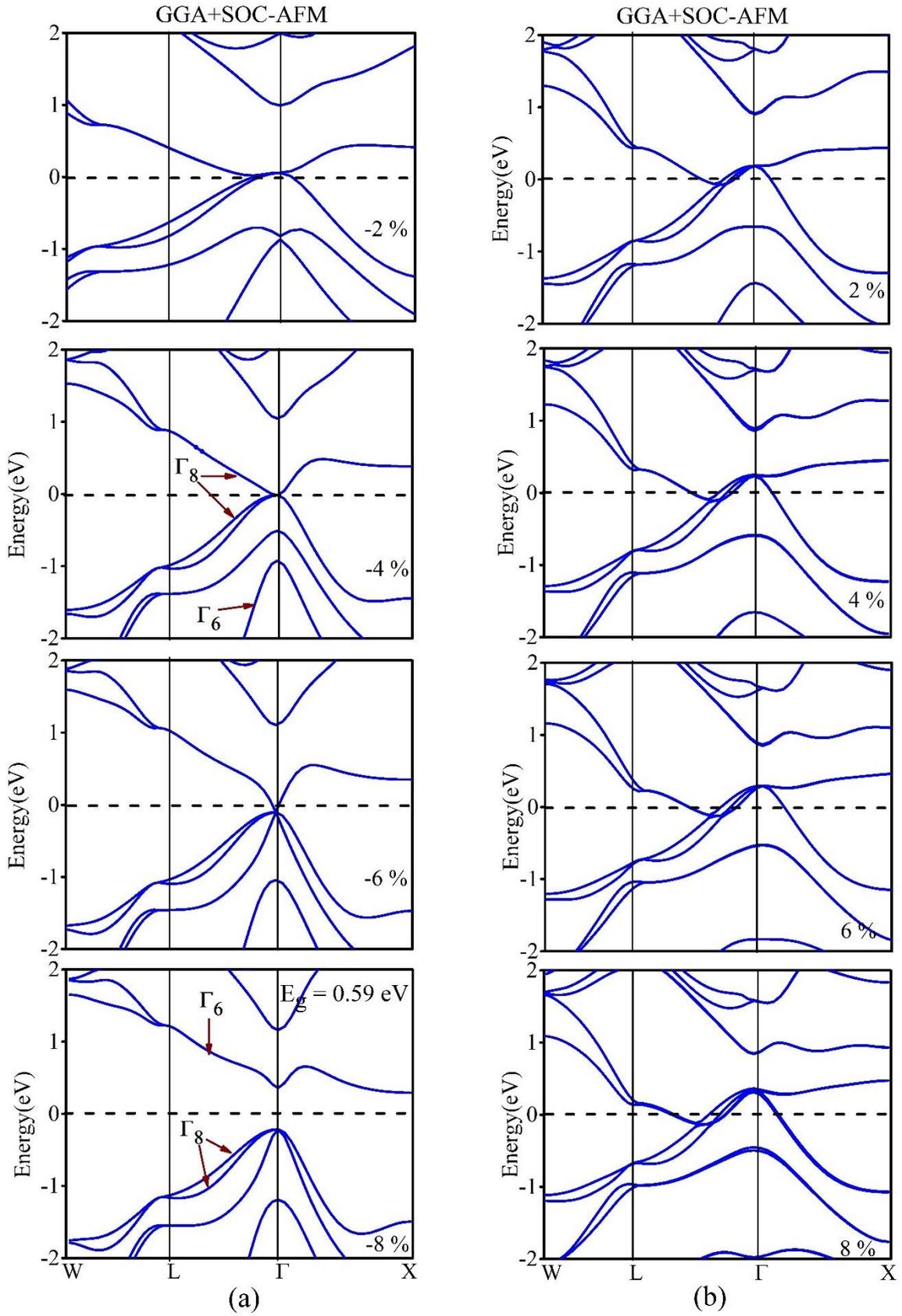

*Figure 6:* Band structures of TbPtBi with SOC under (a) compressive and (b) tensile strain.

## 4. Discussion

In most semiconductors, *s*-type orbitals have higher energy than *p*-type orbitals and are located in the conduction band, whereas *p*-type orbitals are located in the valence band. However, in the presence of strong spin-orbit coupling in the heavy elemental compound, this picture is inverted. Where the *p*-type orbital becomes in the higher energy states, and the band inversion phenomenon occurs. This band inversion is inextricably linked to the topology [31]. Following the pioneering work of Fu and Kene [32] and Xiao et al. [27], the energy difference between the energies of s-type and p-type states can be used to characterize the topology of the electronic band structure. If $\Delta E$ is positive in the HH compound, it represents the topologically non-trivial state, while a negative value indicates the topologically trivial state. Our findings convincingly demonstrate that the system TbPtBi has a topologically semi-metallic phase. However, when subjected to compressive strain, its non-trivial character vanishes. The question is, why does it happen in this manner? We argued that the effects of spin-orbit coupling and compressive strain are diametrically opposed. SOC creates *p*-type orbitals in the higher energy state and shifts them over *s*-type orbitals (band inversion), whereas compressive strain does the reverse.

To comprehend this process, we must grasp that it is the change in hybridization strength caused by a change in the lattice constant (lattice constant reduces because of the compressive strain). For p-type and s-type bands, the change is substantially different. Lobs (in the x-y direction) in p-type orbitals make this transition far more pronounced than in spherically symmetric s-type orbitals. Furthermore, hybridization with a lower lattice constant results in lower energy p-type orbitals. This is why the *s*-type and *p*-type bands flip over when the lattice constant changes (compressive strain). Our study predicts the importance of external strain as a centered parameter.

## 5. Conclusion

Based on first principle calculations, we have systematically investigated various parametric phase transitions of the TbPtBi compound without and with the effect of SOC and both compressive and tensile strain. The TbPtBi compound undergoes a phase transition from metal to semiconductor when compressed by $-4\%$ to $-6\%$. The narrow band gap of $0.16\mathrm{eV}$ is opened with a strong compressive strain of $-6\%$. When compressive strain is applied along with the SOC, the TbPtBi compound follows two phase transitions with band inversion characteristics, (i) from topological semimetal to trivial semimetal and (ii) from trivial semimetal to semiconductor just by increasing the degree of strain. External contractibility via strain makes this compound a potential candidate for strain-controllable thermoelectric devices.

## 6. Acknowledgment